\documentclass[11pt]{article}
\usepackage{amsthm,amssymb,amsfonts,amsmath,amscd}
\usepackage{nccmath}
\usepackage{graphicx}
\usepackage{hyperref}

\makeatletter \@addtoreset{equation}{section} \makeatother

\newtheorem{theorem}{Theorem}[section]

\newtheorem{remark}{Remark}[section]
\newtheorem{definition}{Definition}[section]

\newcommand{\mdet}{\mathrm{det}}

\newcommand{\Tr}{\mathrm{Tr}\,}
\newcommand{\Str}{\mathrm{Str}\,}

\begin{document}

\title{Transfer operator approach to 1d random band matrices}
\author{ Mariya Shcherbina
\thanks{Institute for Low Temperature Physics, Kharkiv, Ukraine\& Karazin Kharkiv National University, Kharkiv, Ukraine,  e-mail: shcherbi@ilt.kharkov.ua} \and
 Tatyana Shcherbina
\thanks{ Department of Mathematics, Princeton University, Princeton, USA, e-mail: tshcherbyna@princeton.edu. Supported in part by NSF grant DMS-1700009.}}

\date{}
\maketitle

\begin{abstract}
We discuss an application of the transfer operator approach  to the analysis
of the different spectral characteristics of 1d random band matrices (correlation functions of
 characteristic polynomials, density of states,  spectral correlation functions). 
 We show that when the bandwidth $W$ crosses the threshold $W=N^{1/2}$, the model
 has a kind of phase transition (crossover), whose nature can be explained
   by the spectral properties of the transfer operator.
\end{abstract}

\section{Introduction}\label{s:1}
Random band matrices (RBM) represent quantum systems on a large box in $\mathbb{Z}^d$ with random transition amplitudes effective up to distances of order $W$, which is called a bandwidth.
They are natural intermediate models to study eigenvalue statistics
and quantum propagation in disordered systems as they interpolate between Wigner
matrices and random Schr$\ddot{\hbox{o}}$dinger operators: Wigner matrix ensembles represent mean-field
models without spatial structure, where the quantum transition rates between any
two sites are i.i.d. random variables; in contrast, random Schr$\ddot{\hbox{o}}$dinger operator has only  a random diagonal potential 
 in addition to the deterministic Laplacian on a box in $\mathbb{Z}^d$. 

In the simplest 1d case RBM  $H$ is a Hermitian or real symmetric
 $N\times N$ matrix with independent (up to the symmetry condition) entries $ H_{ij}$ 
 such that 
 \[\mathbb{E}\big\{ H_{ij}\}=0, \quad\mathbb{E}\{\,|H_{ij}|^2\}=(2W)^{-1}\mathbf{1}_{|i-j|\le W},\]
 i.e. $H$ is a Hermitian matrix with which has inly non  $2W+1$ zero diagonals whose entries are i.i.d. random variables (up to the symmetry)
 and the sum of the variances of entries in each line is $1$.
 
In a more general case $H$ is a Hermitian  random 
$N\times N$ matrix,  whose entries $H_{jk}$ are independent (up to the symmetry) complex  random
variables with mean zero and variances scaled as
\begin{equation}\label{H_gen}
\mathbb{E}\{|H_{jk}|^2\}=\dfrac{1}{W^d}J\Big(\dfrac{|j-k|}{W}\Big).
\end{equation}
Here $\Lambda$ is a box in $\mathbb{Z}^d$, $|\Lambda|=N$, and $J: \mathbb{R}^d\to \mathbb{R}_+$ is a function having the compact support or decaying sufficiently fast at infinity and
normalized in such a way that 
\[
\sum\limits_{k\in\Lambda} J(|k|/W) =1,
\]
and the bandwidth $W\gg 1$ is a large parameter. 
 
 The density of states $\rho$ of a general class of RBM with $W\gg  1$ is given by the well-known Wigner semicircle law (see
\cite{BMP:91, MPK:92}):
\begin{equation}\label{rho_sc}
\rho(E)=(2\pi)^{-1}\sqrt{4-E^2},\quad E\in[-2,2].
\end{equation}
As it was mentioned above, a substantial interest to random band matrices is caused by the fact that
they have a non-trivial spatial structure like random
Schr$\ddot{\hbox{o}}$dinger matrices (in contrast to classical random matrix ensembles),
and furthermore RBM and random
Schr$\ddot{\hbox{o}}$dinger matrices are expected to have some similar qualitative
properties (for more details on these conjectures see \cite{Sp:12}).
For instance, RBM can be used
to model the celebrated Anderson metal-insulator phase transition in $d\ge 3$. Moreover,
the crossover for RBM can be investigated even in $d = 1$ by varying the bandwidth $W$.

The key physical parameter of RBM  is the localization length $\ell_\psi$,
which describes the length scale of the eigenvector $\psi(E)$ corresponding to the energy $E\in (-2,2)$.
The system is called delocalized if for all $E$ in the bulk of spectrum $\ell_\psi$  is
comparable with the system size, $\ell_\psi\sim N$, and it is called localized otherwise.
Delocalized systems correspond to electric
conductors, and localized systems are insulators.
 
 In the case of 1d RBM there is a fundamental conjecture stating that for every  eigenfunction $\psi(E)$ in the bulk of the
 spectrum  $\ell_\psi$ is
of order $W^2$ (see \cite{Ca-Co:90, FM:91}). In $d=2$, the localization length is expected to be exponentially large in $W$, 
in $d\ge 3$ it is expected to be macroscopic, $\ell_\psi\sim N$, i.e. system is delocalized.

Notice that the global eigenvalue statistics for 1d RBM such as density of states does not feel any difference between the regime $W\gg  \sqrt N$ 
and $W\ll \sqrt N$ (see (\ref{rho_sc})).
Same situation with the 
 central limit theorem for the linear eigenvalue statistics which was proved in \cite{Sh:15}  for any $W\gg  1$ (see also
\cite{So:13} for CLT in the regime $W\gg  \sqrt N$). However, the questions of the localization length  are closely related to the universality conjecture
of the bulk \textit{local} regime of the random matrix theory. The bulk local regime deals with the behaviour of eigenvalues of $N\times N$
random matrices on the intervals whose length is of the order $O(N^{-1})$.
According to the Wigner -- Dyson universality conjecture, this local behaviour does not depend on the matrix probability
law (ensemble) and is determined only by the symmetry type of matrices (real
symmetric, Hermitian, or quaternion real in the case of real eigenvalues and orthogonal,
unitary or symplectic in the case of eigenvalues on the unit circle).
In terms of eigenvalue statistics
the conjecture about the localization length of RBM in $d=1$ means that  1d RBM in the bulk of the spectrum changes the spectral local behaviour of random operator type with
Poisson local eigenvalue statistics (for $W\ll \sqrt{N}$) to the local
spectral behaviour of the GUE/GOE type (for $W\gg  \sqrt{N}$). In particular, it means that if
we consider the second correlation function $R_{2}$ defined by the equality
\begin{equation}\label{R_2} 
\mathbb{E}\Big\{ \sum_{j_{1}\neq  j_{2}}\varphi
(\lambda_{j_{1}},\lambda_{j_{2}})\Big\}=\int_{\mathbb{R}^{2}} \varphi
(\lambda_{1},\lambda_{2})R_{2}(\lambda_{1},\lambda_{2})
d\lambda_{1} d\lambda_{2},
\end{equation}
where $\{\lambda_j\}$ are eigenvalues of a random matrix, the function $\varphi: \mathbb{R}^{2}\rightarrow \mathbb{C}$ is
bounded, continuous and symmetric in its arguments, and the
summation is over all pairs of distinct integers $
j_{1},j_{2}\in\Lambda$, 
then in the delocalization  region $W\gg \sqrt{N}$
\begin{equation}\label{Un}
(N\rho(E))^{-2}
R_2\left(E+\displaystyle\frac{\xi_1}{\rho(E)\,N},
E+\displaystyle\frac{\xi_2}{\rho(E)\,N}\right)\longrightarrow
1-\dfrac{\sin^2 (\pi(\xi_1-\xi_2))}
{\pi^2(\xi_1-\xi_2)^2},
\end{equation}
while in the localization region
\begin{equation}\label{deloc}
(N\rho(E))^{-2}
R_2\left(E+\displaystyle\frac{\xi_1}{\rho(E)\,N},
E+\displaystyle\frac{\xi_2}{\rho(E)\,N}\right)\longrightarrow
1.
\end{equation}
The conjecture on the crossover in RBM with $W\sim\sqrt N$ is supported by physical derivation due to Fyodorov and Mirlin (see \cite{FM:91}) based on supersymmetric formalism, 
and also by the so-called Thouless scaling. However, there are only a few partial results on the mathematical level of rigour.
At the present time only some upper and lower bounds for $\ell_\psi$ for the general class of 1d RBM are proved rigorously. 
It is known from the paper \cite{S:09} that $\ell_\psi\le W^8$. Recently this bound was improved in \cite{wegb:16} to $W^7$.
On the other side, for the general Wigner matrices (i.e. $W=N$) the
bulk universality has been proved in \cite{EYY:10, TV:11}, which gives $\ell_\psi \ge W$. 
By the developing the Erd\H{o}s-Yau approach,  there were also obtained some other results, where the localization length is controlled: $\ell_\psi \ge W^{7/6}$ in \cite{EK:11} and $\ell_\psi\ge W^{5/4}$ in \cite{Yau:12}.
GUE/GOE gap distributions for $W\sim N$ was proved recently in \cite{BEYY:16}.

The study of the   eigenfunctions decay is closely related to properties of the Green function $(H-E-i\varepsilon)^{-1}$
with a small $\varepsilon$. For instance, if $|(H-E-i\varepsilon)^{-1}_{ii}|^2$ (without expectation) is bounded for all $i$
and all $E\in (-2,2)$,  then each normalized eigenvector $\psi$ of $H$ is delocalized on the scale $\varepsilon^{-1}$ in a sense that
\[
\max_i |\psi_i|^2\le C \varepsilon^{-1},
\]
and so $\psi$ is supported on at least $\varepsilon^{-1}$ sites. In particular, if $|(H-E-i\varepsilon)^{-1}_{ii}|^2$ can be controlled down to
the scale $\varepsilon\sim 1/N$, then the system is in the complete delocalized regime.
Moreover, in view of the bound
\[
\mathbb{E}\{|(H-E-i\varepsilon)^{-1}_{jk}|^2\}\sim C\varepsilon^{-1}\, e^{-\|j-k\|/\ell}
\]
which holds for the localized regime, the problem of localization/delocalization reduces to controlling
\[
\mathbb{E}\{|(H-E-i\varepsilon)^{-1}_{jk}|^2\}
\]
for $\varepsilon\sim 1/N$. As it will be shown below, similar estimates  of $\mathbb{E}\{|\Tr (H-E-i\varepsilon)^{-1}|^2\}$ for $\varepsilon\sim 1/N$ 
are required to work with the correlation functions of RBM.

Despite many attempts, such control so far has not been achieved. The standard approaches of \cite{EYY:10} and \cite{Yau:12} do 
not seem to work for $\varepsilon \le W^{-1}$, and so cannot give an information about the strong form of delocalization (i.e. for \textit{all} eigenfunctions). 
Classical moment methods, even with a delicate
renormalization approach \cite{S:11}, could not break the barrier $\varepsilon\sim W^{-1}$ either.

Another method which  
allows to work with random operators with non-trivial spatial structures and breaks that barrier, is supersymmetry techniques (SUSY). 
It is based on the representation of the determinant as an integral (formal) over the Grassmann variables. 
Combining this representation with the representation of the inverse determinant as an integral over the Gaussian complex field,
SUSY allows to obtain the  integral representation for the main spectral characteristics such as averaged density of states and 
correlation functions, as well as for 
$\mathbb{E}\{G_{jk}(E+i\varepsilon)\}$, $\mathbb{E}\{|G_{jk}(E+i\varepsilon)|^2\}$, etc.
For instance, according to the properties of  the Stieljes transform, the second correlation function
can be rewritten in the form
\begin{align}\label{cor=det}
R_{2}(\lambda_1,\lambda_2)=&(\pi N)^{-2}\lim_{\varepsilon\to 0}
\mathbb{E}\{\Im\,\Tr(H-\lambda_1-i\varepsilon)^{-1}\Im\, \Tr (H-\lambda_2-i\varepsilon)^{-1}\}\\ \notag
=&(2i\pi N)^{-2}\lim_{\varepsilon\to 0}
\mathbb{E}\Big\{\Big(\Tr(H-\lambda_1-i\varepsilon)^{-1}-\Tr(H-\lambda_1+i\varepsilon)^{-1}\Big)\\ \notag
&\qquad\qquad\qquad\qquad\quad\quad\times \Big(\Tr(H-\lambda_2-i\varepsilon)^{-1}-\Tr(H-\lambda_2+i\varepsilon)^{-1}
\Big)\Big\},
\end{align}
and since
\begin{align}\label{r_det}
&\mathbb{E}\{\Tr (H-z_1)^{-1}\Tr (H-z_2)^{-1}\}= \frac{d^2}{dz_1'dz_2'}\mathbb{E}\Big\{\frac{ \det(H-z_1)\det(H-z_2))}{\det(H-z_1')\det(H-z_2'))}\Big\}\Big|_{z'=z},
\end{align}
$R_{2}$ can be represented as a sum of derivatives of the expectation of ratio of four determinants. Besides, it is expected that if we set 
 \begin{align}\label{z}
 z_1&=E+i\varepsilon/N+\xi_1/N\rho(E),\quad z_2=
E+i\varepsilon/N+\xi_2/N\rho(E),\\ \notag
z_1^\prime&=E+i\varepsilon/N+\xi_1^\prime/N\rho(E),\quad z_2^\prime=
E+i\varepsilon/N+\xi_2^\prime/N\rho(E),
\end{align}
then the r.h.s. of (\ref{r_det})  before taking derivatives is an analytic function in $\xi_1,\xi_2,\xi_1',\xi_2'$. Thus, to study the second correlation function,
it suffices to study the ratio of four determinants, which we call the second "generalized" correlation functions
\begin{align}\label{calR_2}
\mathcal{R}_{2}^{+-}(z_1,z_1';z_2, z_2')&=\mathbb{E}\bigg\{\dfrac{\mdet(H-z_1)\mdet(H-\overline{z}_2)}
{\mdet(H-z_1')\mdet(H-\overline{z}_2^\prime)}\bigg\},\\ \notag
\mathcal{R}_{2}^{++}(z_1,z_1';z_2, z_2')&=\mathbb{E}\bigg\{\dfrac{\mdet(H-z_1)\mdet(H-z_2)}
{\mdet(H-z_1')\mdet(H-z_2^\prime)}\bigg\}.
\end{align}
 Similarly the derivative of the first "generalized" correlation function
\begin{align*}
&\mathcal{R}_1(z_1,z_1'):=\mathbb{E}\Big\{\frac{ \det(H-z_1')}{\det(H-z_1)}\Big\}
\end{align*}
 gives the Stieltjes transform of of the density of states (the first correlation function). 
 
 Instead of eigenvalue correlation functions one can consider more simple objects which are  the
correlation functions of characteristic polynomials:
\begin{equation}\label{R_0}
\mathcal{R}_0(\lambda_1,\lambda_2)=\mathbb{E}\Big\{ \det(H-\lambda_1)\det(H-\lambda_{2})\Big\},
\quad \lambda_{1,2}=E\pm\xi/N\rho(E).
\end{equation}
Characteristic polynomials are the objects of independent interest because of their connections to the
number theory, quantum chaos, integrable systems, combinatorics, representation
theory and others. But in our context the main point is that
from the SUSY point of view correlation functions of characteristic polynomials correspond to the so-called fermion-fermion (Grassmann) sector of the supersymmetric full model
describing the usual correlation functions (since they represent two determinants in the numerator of (\ref{calR_2})). They are especially convenient 
for the SUSY approach and were successfully studied by the techniques
for many ensembles (see \cite{Br-Hi:00}, \cite{Br-Hi:01}, \cite{TSh:11}, \cite{TSh:11_1}, etc.).
In addition, although $\mathcal{R}_{0}(\lambda_1,\lambda_2)$ is not a local object, it is also expected 
to be universal in some sense. Moreover, correlation functions of characteristic polynomials are expected to exhibit a crossover which is similar to that
of local eigenvalue statistics. In particular, for  1d RBM they are expected to have the same local behaviour 
as for GUE for $W\gg  \sqrt{N}$, and the different behaviour for $W\ll \sqrt{N}$. Besides, the analysis of $\mathcal{R}_{0}(\lambda_1,\lambda_2)$
is much less involved than that for $\mathcal{R}_{2}^{+-}(z_1,z_1';z_2, z_2')$, but on the other hand, this analysis
 allows to understand the nature of the crossover in RBM when $W$ crosses the threshold 
$W\sim \sqrt{N}$. 

The derivation of SUSY integral representation is basically an algebraic step, and usually it can be done by the standard algebraic manipulations.
SUSY is widely  used in
the physics literature, but the rigour analysis of the obtained integral representation is a real mathematical challenge. Usually it is quite difficult, and it requires a powerful 
analytic and statistical mechanics techniques, such as a saddle point analysis, transfer operators, cluster expansions, renormalization
group methods, etc. However, it can be done rigorously for some special class of RBM. 

There exist especially convenient classes of RBM, where the control of SUSY integral representation becomes more accessible.
One of them was introduced in \cite{DPS:02}: it is (\ref{H_gen}) with Gaussian elements with variance
\begin{equation}\label{Jb}
\mathbb{E}\{|H_{jk}|^2\}=\left(-W^2\Delta+1\right)^{-1}_{jk},
\end{equation}
where $\triangle$ is the discrete Laplacian on $\Lambda$ with Neumann boundary conditions: for the case $d=1$,
\begin{equation}\label{lapl}
(-\Delta f)_j=\left\{\begin{array}{ll} -f_{j-1}+2f_j-f_{j+1},&j\ne 1,n,\\
-f_{j-1}+f_j-f_{j+1},& j=1,n
\end{array}\right.
\end{equation}
with $f_{0}=f_{n+1}=0$. It is easy to see that in 1d case 
$J_{jk}\approx C_1W^{-1}\exp\{-C_2|j-k|/W\}$, and so the variance of matrix
elements is exponentially small when $|j-k|\gg W$.

Another class of convenient models are the Gaussian block RBM which are the special class of Wegner's orbital models (see \cite{We:79}). 
Gaussian block RBM are $N\times N$ Hermitian block matrices composed from $n^2$ blocks of the size $W\times W$ ($N=nW$).
Only 3 block diagonals are non zero:
\begin{align*}
H=\left(\begin{array}{ccccccc}
A_1&B_1&0&0&0&\dots&0\\
B_1^*&A_2&B_2&0&0&\dots&0\\
0&B_2^*&A_3&B_3&0&\dots&0\\
.&.&B_3^*&.&.&.&.\\
.&.&.&.&.&A_{n-1}&B_{n-1}\\
0&.&.&.&0&B_{n-1}^*&A_n
\end{array}\right).
\end{align*}
Here
$A_1,\dots A_n$ are independent $W\times W$ $GUE$-matrices with i.i.d. (up to the symmetry) Gaussian entries with variance $(1-2\alpha)/W$,
$\quad\alpha<\frac{1}{4}$, and
$B_1,\dots B_{n-1}$  are independent $W\times W$ Ginibre matrices with  i.i.d. Gaussian entries with variance $\alpha/W$.

More precisely, $H$ is
Hermitian matrices  with complex zero-mean random Gaussian entries $H_{jk,\alpha\beta}$,
where $j,k \in\Lambda\subset \mathbb{Z}^d$ (they parameterize the lattice sites) and $\alpha, \gamma= 1,\ldots, W$ (they
parametrize the orbitals on each site), such that
\begin{equation}\label{H}
\langle H_{j_1k_1,\alpha_1\gamma_1}H_{j_2k_2,\alpha_2\gamma_2}\rangle=\delta_{j_1k_2}\delta_{j_2k_1}
\delta_{\alpha_1\gamma_2}\delta_{\gamma_1\alpha_2} J_{j_1k_1}
\end{equation}
with
\begin{equation}\label{J_old}
J=1/W+\alpha\Delta/W,
\end{equation}
where $W\gg  1$ and $\Delta$ is the discrete Laplacian on $\Lambda$ (as in (\ref{Jb})). 
The probability law of $H$ can be written in the form
\begin{equation}\label{pr_l}
P_N(d H)=\exp\Big\{-\dfrac{1}{2}\sum\limits_{j,k\in\Lambda}\sum\limits_{\alpha,\gamma=1}^W
\dfrac{|H_{jk,\alpha\gamma}|^2}{J_{jk}}\Big\}dH.
\end{equation}
This model is one of the possible realizations of the Gaussian RBM, for example for
$d=1$ they correspond to the band matrices with the bandwidth  $2W+1$.
Let us remark that for this model $N=nW$, hence the crossover is expected for $n\sim W$.

The main advantage of both models (\ref{Jb}) and (\ref{H}) -- (\ref{J_old}) is that the main spectral
 characteristics such as density of states, $R_{2}$, $\mathbb{E}\{|G_{jk}(E+i\varepsilon)|^2\}$ for these models can
 be expressed via SUSY as the averages of certain observables of  \textit{nearest-neighbour} statistical mechanics models on $\Lambda$, which makes
 the model easier. For instance,  the detailed information about the averaged density of states Gaussian RBM (\ref{Jb}) in dimension 3 including local semicircle low
at arbitrary short scales  and smoothness in 
energy (in the limit of infinite volume and fixed large band width $W$) was obtained in \cite{DPS:02}. The techniques of this paper was used
in \cite{DL:16} to obtain the same result in 2d. The rigorous application of SUSY to the Gaussian block RBM (\ref{H}) -- (\ref{J_old}) 
was developed in \cite{TSh:14_1}, where the universality of the bulk local regime for $n=const$ was proved.
Combining this approach with Green's function comparison strategy it has been proved in \cite{EB:15} that  $\ell \ge W^{7/6}$
(in a strong sense) for the block band matrices with rather general element's distribution. 

The nearest-neighbour structure of the model also
allows to combine the SUSY techniques with a transfer matrix approach.

\section{Idea of the transfer operator approach}\label{s:2}
The supersymmetric transfer matrix formalism 
 was first suggested by Efetov (see \cite{Ef}) and on a heuristic level it was adapted specifically for RBM
in \cite{FM:94} (see also references therein). The rigorous application of this method
to the density of states and correlation function of characteristic polynomials was done in \cite{SS:den}, \cite{SS:ChP}, \cite{SS:sigma}, \cite{TSh:17}.
The approach is based on the fact that many
nearest-neighbour statistical mechanics problems in 1d can be formulated in terms of
properties of some integral operator $K$ that is called a transfer operator. 
 More precisely, the discussion above yields that for 1d RBM of the form (\ref{Jb}) or (\ref{H}) -- (\ref{J_old})  the SUSY techniques helps
 to find a scalar kernel $\mathcal{K}_0(X_1,X_2)$ and matrix kernels $\mathcal{K}_1(X_1,X_2)$, $\mathcal{K}_2(X_1,X_2)$ (containing
 $z_{1,2},z_{1,2}'$ as parameters)   such that
\begin{align}\label{int_rep}
&\mathcal{R}_0(\lambda_1,\lambda_2)=C_N\int g_0(X_1)\mathcal{K}_0(X_1,X_2)\dots \mathcal{K}_0(X_{n-1},X_n)f_0(X_n)  \prod d X_i,\\
&\mathcal{R}_1(z_1,z_1')=W^2\int g_1(X_1)\mathcal{K}_1(X_1,X_2)\dots \mathcal{K}_1(X_{n-1},X_n)f_1(X_n)  \prod d X_i,\notag\\
&\mathcal{R}_2(z_1,z_1';z_2,z_2')=W^4\int g_2(X_1)\mathcal{K}_2(X_1,X_2)\dots \mathcal{K}_2(X_{n-1},X_n)f_2(X_n)  \prod d X_i,
\notag\end{align}
where $\{X_j\}$ are Hermitian $2\times 2$ matrices for the cases of $\mathcal{R}_{0}$,  $2\times 2$ matrices whose
entries depend on 2 spacial variables $x_{1j},y_{1j}\in \mathbb{R}$ for the cases $\mathcal{R}_{1}$,  and  for the case of $\mathcal{R}_2$
 $\{X_j\}$ are  $70\times 70$ matrices whose
entries depend on 4 spacial variables $x_{1j},x_{2j},y_{1j},y_{2j}\in \mathbb{R}$, unitary $2\times 2$ matrix $U_i$,
 and hyperbolic $2\times 2$ matrix $S_j$,
 $dX_j$ means the standard measure on $\hbox{Herm}(2)$ for $\mathcal{R}_{0}$,  $dX_j=dx_{j1}d y_{j1}$ for $\mathcal{R}_1$, and  for $\mathcal{R}_2$ $dX_j$ means the integration over 
 $dx_{1j}dx_{2j}dy_{1j}dy_{2j}dU_jdS_j$ with $dU, dS$ being the
corresponding Haar measures.

Remark, that for the model (\ref{Jb}) $n=N$,  while for the block band matrix (\ref{H}) -- (\ref{pr_l}) $n$ is a number of blocks on the main diagonal.

  The idea of the transfer operator approach is very simple and natural.
 Let $\mathcal{K}(X,Y)$ be the matrix kernel of the compact  integral operator in $\oplus_{i=1}^pL_2[X,d\mu(X)]$. Then 
\begin{align}\notag
&\int g(X_1) \mathcal{K}(X_1,X_2)\dots \mathcal{K}(X_{n-1},X_n)f(X_n)   \prod d\mu(X_i)=(\mathcal{K}^{n-1}f,\bar g)\\
&=\sum_{j=0}^\infty\lambda_j^{n-1}(\mathcal{K})c_j,\quad with\quad
c_j=(f,\psi_j)(g,\tilde\psi_j).
\label{2}
\end{align}
Here $\{\lambda_j(\mathcal{K})\}_{j=0}^\infty$ are the eigenvalues of $\mathcal{K}$ ( $|\lambda_{0}|\ge |\lambda_{1}|\ge\dots$),
$\psi_j$ are corresponding eigenvectors, and $\tilde \psi_j$ are the eigenvectors of $\mathcal{K}^*$. Hence, to study the correlation function,
it suffices to study the eigenvalues and eigenfunctions of the integral operator  with a kernel $\mathcal{K}(X,Y)$.

The main difficulties here are
the complicated structure and non self-adjointness of the corresponding transfer operators. 

In fact, since the analysis of eigenvectors of  non self-adjoint operators is rather involved, it is  simpler  to work with the resolvent 
analog of (\ref{2})
\begin{align}\label{res_rep}
\mathcal{R}_{\alpha}=(\mathcal{K}_\alpha^{n-1}f,\bar g)=-\frac{1}{2\pi i}\oint_{\mathcal{L}}z^{n-1}(\mathcal{G}_\alpha(z)f,\bar g)dz,\quad 
\mathcal{G}_\alpha(z)=(\mathcal{K}_\alpha-z)^{-1},
\quad \alpha=0,1,2,
\end{align}
where $\mathcal{L}$ is any closed contour which contains all eigenvalues of $\mathcal{K}_\alpha$. 
For any $\alpha$ if we set
\[\lambda_*=\lambda_0(\mathcal{K}_\alpha),\quad(\lambda_*\sim 1),\]
then it suffices to choose $\mathcal{L}$ as $\mathcal{L}_0=\{z:|z|=|\lambda_*|(1+O(n^{-1}))\}$. However, it is more 
convenient to
 choose $\mathcal{L}=\mathcal{L}_1\cup \mathcal{L}_2$, where  $\mathcal{L}_2=\{z:|z|=|\lambda_*|(1-\log^2n/n)\}$, and $\mathcal{L}_1$ is some contour in the domain between
 $\mathcal{L}_0$ and $\mathcal{L}_2$ which contains all eigenvalues of $\mathcal{K}_\alpha$ outside of $\mathcal{L}_2$. Then
\begin{align*}
(\mathcal{K}_\alpha^{n-1}f,\bar g)=-\frac{1}{2\pi i}\oint_{\mathcal{L}_1} z^{n-1}(\mathcal{G}_\alpha(z)f,\bar g)dz-\frac{1}{2\pi i}\oint_{\mathcal{L}_2} 
z^{n-1}(\mathcal{G}_\alpha(z)f,\bar g)dz
\end{align*}
and if we have a reasonable bound for $\|\mathcal{G}_\alpha(z)\|$ ($z\in \mathcal{L}_2$), then the second integral is small comparing with $|\lambda_*|^{n-1}$, since 
\[|z|^{n-1}\le|\lambda_*|^{n-1}e^{-\log^2n}. \]
Hence, it is natural to expect that the integral over $\mathcal{L}_1$ gives the main contribution to $\mathcal{R}_{\alpha}$.
\begin{definition}\label{def:1}
 We shall say that the operator $\mathcal{A}_{n,W}$  is equivalent to  $\mathcal{B}_{n,W}$  ($\mathcal{A}_{n,W}\sim\mathcal{B}_{n,W}$),
 if for some certain   contour $\mathcal{L}_1$
(the choice of $\mathcal{L}_1$ depends on the problem)
\[ ((\mathcal{A}_{n,W}-z)^{-1}f,\bar g)= ((\mathcal{B}_{n,W}-z)^{-1}f,\bar g)(1+o(1)),\quad n,W\to \infty,\]
with $f,g$ of (\ref{2}).
\end{definition}
The idea  is to find  some $\mathcal{K}_{*\alpha}\sim \mathcal{K}_\alpha$  whose spectral analysis we are ready to perform.

 \section{Mechanism of the crossover for $\mathcal{R}_0$}\label{s:3}
As it was mentioned in Section \ref{s:1}, the simplest object which allows to understand the crossover's mechanism  for the 1d RBM 
(\ref{Jb}) is the correlation function of characteristic polynomials  $\mathcal{R}_0$.
Using SUSY and the idea of the transfer operator approach, one can write $\mathcal{R}_0$ (see \cite{SS:ChP}) as
\begin{align}\label{F_rep}
\mathcal{R}_0\Big(E+\dfrac{\xi}{N\rho(E)},E-\dfrac{\xi}{N\rho(E)}\Big)=C_n\cdot W^{-4n}\mdet^{-2} J\cdot ( K^{n-1}_{0\xi}\mathcal{F}_{\xi},\bar{\mathcal{F}}_{\xi}),
\end{align}
where $(\cdot,\cdot)$ is a standard inner product in $L_2( \hbox{Herm}(2),dX)$ (i.e., $2\times 2$ Hermitian matrices), 
with respect to the measure 
\begin{equation*}
dX_j=d(X_{j})_{11}d(X_{j})_{22}d\Re (X_{j})_{12}d\Im (X_{j})_{12},
\end{equation*}
$C_n$ is some $\xi$-independent constant,
 $ K_{0\xi}: \mathcal{H}\to\mathcal{H}$ be the operators with the kernels 
\begin{align} \label{K}
K_{\xi}(X,Y)&=\dfrac{W^4}{2\pi^2}\,\mathcal{F}_\xi(X)\,\exp\Big\{-\frac{W^2}{2}\Tr
(X-Y)^2\Big\}\,\mathcal{F}_\xi(Y).
\end{align}
where $\hat\xi=\hbox{diag}\,\{\xi,-\xi\}$, $\Lambda_0=E\cdot I_2$,  and
 $\mathcal{F}_\xi (X)$ is the operator of multiplication  by
\begin{align}\label{F_cal}
\mathcal{F}_\xi (X)&=\mathcal{F} (X)\cdot \exp\Big\{-
\frac{i}{2n\rho(E)}\, \Tr X\hat\xi\Big\} 
\end{align}
with
\[
 \mathcal{F}(X)=\exp\Big\{-
\frac{1}{4}\, \Tr\Big(X+\frac{i\Lambda_0}{2}\Big)^2+\frac{1}{2}\,\Tr \log
\big(X-i\Lambda_0/2\big)-C_+\Big\}
\]
and some specific $C_+$. Notice that the stationary points of $\mathcal{F}$ are
\begin{equation}\label{a_pm}
a_+=-a_-= \sqrt{1-E^2/4}=\pi\rho(E).
\end{equation}

The first step is to show that if we introduce the projection $P_{\pm}$  onto  the $W^{-1/2}\log W$-neighbourhood of the``surface" $X_*(U)=UDU^*$ with   $D=\hbox{diag}\,\{a_+,a_-\}$ 
and $U\in \mathring{U}(2):=U(2)/U(1)\times U(1)$, then in the sense of Definition \ref{def:1}
\begin{align}\label{R_0_rep}
K_{0\xi}\sim P_{\pm}K_{0\xi}P_{\pm}.
\end{align}

To study the operators $ P_{\pm}K_{0\xi}P_{\pm}$  we  use the  "polar coordinates". Namely,
introduce
\begin{align}\label{t}
t=(x_1-y_1)(x_2-y_2), \quad p(x,y)=\dfrac{\pi}{2}(x-y)^2,
\end{align}  
and denote by $dU$ the integration with respect to the Haar measure
on the group $\mathring{U}(2)$. Consider the space $L_2[\mathbb{R}^2,p]\times L_2[\mathring{U}(2),dU]$.
The inner product and the action of an integral operator in this space are
\begin{align}\label{(,)_a}
&(f,g)_p=\int f(x,y)\bar g(x,y)p(x,y)\,dx\,dy;\\
&(Mf)(x_1,y_1,U_1)=\int M(x_1,y_1,U_1;x_2,y_2,U_2)\,f(x_2,y_2,U_2)\,p(x_2,y_2)dx_2dy_2dU_2.
\notag\end{align} 
Changing the variables
\[X=U^*\Lambda U,\quad \Lambda=\mathrm{diag}\{x_1,x_2\},\quad x_1>x_2,\quad U\in \mathring{U}(2),\]
we obtain that $K_{0\xi}$ can be  represented as an integral operator in $L_2[\mathbb{R}^2,p]\times L_2[\mathring{U}(2),dU]$  defined by the kernel
\begin{align}\label{rep_2}
&\mathcal{K}_{0\xi}(X,Y)\to \mathcal{K}_{0\xi}(x_1,y_1,U_1;x_2,y_2,U_2)
\end{align}
where
\begin{align} \notag
&\mathcal{K}_{0\xi}(x_1,y_1,U_1;x_2,y_2,U_2)=t^{-1}A_1(x_1,x_2)A_2(y_1,y_2)K_{*0\xi}(t,U_1,U_2)(1+O(n^{-1}W^{-1/2}));\\ \label{A}
&A_{1,2}(x_1,x_2)=(2\pi)^{-1/2}e^{-W^2(x_1-x_2)^2/2}e^{f_{1,2}(x_1)+f_{1,2}(x_2)};\\ \label{K(U)}
&K_{*0\xi}(t,U_1,U_2):=W^2t\cdot e^{tW^2\Tr U_1U_2^*L(U_1U_2^*)^*L/4-tW^2/2}e^{-i\xi\pi(\nu(U_1)+\nu(U_2))/n};\\
&\nu(U)=\Tr U^*LUL/2, \quad L=\mathrm{diag}\{1,-1\},
\notag\end{align}
and $t$ is defined in (\ref{t}). The concrete form of $f_{1,2}$ in (\ref{A}) is not important for us now. It is important that they are analytic functions with  stationary points
$a_{\pm}$ (see (\ref{a_pm})). The analysis of the resolvent of $A_1$ and $A_2$ allows us to show that only 
eigenfunctions localized in the $W^{-1/2}\log W$ neighbourhood of $a_+$ and $a_-$ give essential contribution
in (\ref{2}). More precisely, the resolvent  analysis of $A_{1,2}$  allows to prove  (\ref{R_0_rep}).
Further resolvent analysis gives 
\begin{align}\label{tens_pr}
&P_{\pm} \mathcal{K}_{0\xi}P_{\pm}\sim\mathcal{K}_{*\xi}\otimes \mathcal{A},\\\notag
&\mathcal{K}_{*\xi}(U_1,U_2):=K_{*0\xi}(t^*,U_1,U_2)\quad with \quad t^*=(a_+-a_-)^2=4\pi^2\rho(E)^2,\\
&\mathcal{A}(x_1,x_2,y_1,y_2)=A_1(x_1,x_2)A_2(y_1,y_2).
\notag
\end{align}
 Then from (\ref{res_rep}) and Definition \ref{def:1} it is easy to obtain
\[\mathcal{R}_{\xi}=C_n(\mathcal{K}_{*\xi}^{n-1}\otimes\mathcal{A}^{n-1}f,\bar g)(1+o(1))=(\mathcal{K}_{*\xi}^{n-1}f_0,f_0)(\mathcal{A}^{n-1}f_1,\bar g_1)(1+o(1)),
\]
where we used that both $f,g$ asymptotically can be replaced by $f_0(U)\otimes f_1(x,y)$ with 
\begin{equation}\label{f0}
f_0\equiv 1.
\end{equation}
If  we introduce
\begin{equation}\label{D_2}
D_2=\mathcal{R}_0(E,E),
\end{equation}
then the above consideration yields
\begin{align}\label{ratio}
D_2^{-1}\mathcal{R}_0\Big(E+\dfrac{\xi}{N\rho(E)},E-\dfrac{\xi}{N\rho(E)}\Big)=\frac{(\mathcal{K}_{*\xi}^{n-1}f_0,f_0)}{(\mathcal{K}_{*0}^{n-1}f_0,f_0)}(1+o(1)).
\end{align}
A good news here is that the operator $\mathcal{K}_{*0}$ is  self-adjoint  and  his kernel depends only on $|(U_1U_2^*)_{12}|^2$. By \cite{Vil:68},
his eigenfunctions are associated Legendre polynomials $P_k^j$. Moreover since $\mathcal{K}_{*0}$ is reduced by the space $\mathcal{E}_0\subset L_2(U(2))$
of the functions which depends only on $|U_{12}|^2$, and $f_0\in\mathcal{E}_0$, we can restrict our spectral analysis to $\mathcal{E}_0$. In this space
eigenfunctions of $\mathcal{K}_{*0}$ are Legendre polynomials $P_j$ and 
it is easy to check that correspondent eigenvalues have
the form
\begin{align}\label{la_j}
\lambda_j=1-j(j+1)/t^*W^2+O((j(j+1)/W^2)^2),\quad j=0,1\dots
\end{align}
with $t^*$ of (\ref{tens_pr}). Moreover, it follows from (\ref{K(U)}) that
\[ \mathcal{K}_{*\xi}=\mathcal{K}_{*0}-2n^{-1}\pi i\xi\hat\nu+o(n^{-1}),\]
where $\hat\nu$ is the operator of multiplication by $\nu$ of (\ref{K(U)}). Thus  the eigenvalues of  $\mathcal{K}_{*\xi}$ are in the $n^{-1}$-neighbourhood of
$\lambda_j$. This implies that for $W^{-2}\gg  n^{-1}=N^{-1}$ 
\[|\lambda_1(\mathcal{K}_{*\xi})|\le 1-O(W^{-2}),\quad\lambda_0=1-2n^{-1}\pi i\xi (\nu f_0,f_0)+o(n^{-1})\]
Since
\[(\nu f_0,f_0)=0,\]
we obtain that the numerator and the denominator of (\ref{ratio}) tends to 1 in this regime.

To study the regime $W^{-2}=Cn^{-1}= CN^{-1}$, observe that the Laplace operator 
$\Delta_U$ on $U(2)$ is also reduced by $\mathcal{E}_0$ and has the same eigenfunctions as $\mathcal{K}_{*0}$ with eigenvalues 
\[\lambda_j^*=j(j+1)\]
Hence, we can write $\mathcal{K}_{*\xi}$ as
\[ \mathcal{K}_{*\xi}\sim 1-n^{-1}(C\Delta_U-2i\xi\pi \nu)\Rightarrow (\mathcal{K}_{*\xi}^{n-1}f_0,f_0)\to (e^{-C\Delta_U+2i\xi\pi\hat\nu}f_0,f_0),
\]
where
\begin{align}\label{Lapl_U}
\Delta_U=-\frac{d}{dx}x(1-x)\frac{d}{dx},\quad x=|U_{12}|^2.
\end{align}
And in the regime $W^{-2}\ll  n^{-1}$  we have $\mathcal{K}_{*0}^{n-1}\to I$ in the strong vector topology,  hence
\[ \mathcal{K}_{*\xi}\sim 1-n^{-1}2i\xi\pi\nu\Rightarrow (\mathcal{K}_{*\xi}^{n-1}f_0,f_0)\to (e^{-2i\xi\pi\hat\nu}f_0,f_0)
\]
and the numerator of (\ref{ratio}) is given by the multiplication of $f_0$ by $e^{-2i\xi\pi\hat\nu}$, which  gives the same form as for the correlation function
of the Wigner model.

The last  result was proved in \cite{TSh:14} with a different method:
\begin{theorem}[\textbf{\cite{TSh:14}}]\label{thm:old}
For the 1d RBM of (\ref{Jb}) with $W^2=N^{1+\theta}$, where $0<\theta\le 1$, we have
\begin{equation}\label{lim}
\lim\limits_{n\to\infty}
D_2^{-1}\mathcal{R}_0\Big(E+\dfrac{\xi}{N\rho(E)},E-\dfrac{\xi}{N\rho(E)}\Big)
=\dfrac{\sin(2\pi\xi)}{2\pi\xi},
\end{equation}
i.e. the limit coincides with that for GUE. The limit is uniform in $\xi$ varying in any compact set $C\subset\mathbb{R}$. Here
$\rho(x)$ and $\mathcal{R}_0$ are defined in (\ref{rho_sc}) and (\ref{R_0}),
$E\in (-2,2)$.
\end{theorem}
The  regime $W^{-2}\gg  N^{-1}$ was studied in \cite{SS:ChP}:
\begin{theorem}\label{thm:new}
For the 1d RBM of (\ref{Jb})  with $1 \ll W\le \sqrt {N /C_*\log N}$ for sufficiently big $C_*$, we have
\begin{equation*}
\lim\limits_{n\to\infty}
D_2^{-1}\mathcal{R}_0\Big(E+\dfrac{\xi}{N\rho(E)}, E-\dfrac{\xi}{N\rho(E)}\Big)
=1,
\end{equation*}
where the limit is uniform in $\xi$ varying in any compact set $C\subset\mathbb{R}$. Here $E\in (-2,2)$, 
and $\rho(x)$, $\mathcal{R}_0$, and $D_2$ are defined in (\ref{rho_sc}),  (\ref{R_0}), and (\ref{D_2}).
\end{theorem}
 
\begin{remark}
Although the result is formulated for $\xi_1=-\xi_2=\xi$ in (\ref{z}), one can prove Theorem \ref{thm:new} for $\xi_1,\xi_2\in [-C,C]\subset \mathbb{R}$ by the
same arguments with minor revisions. The only difference is a little bit more complicated expressions for $D_2$ and  $K_\xi$.
\end{remark}
The regime $W^{-2}= C_*N^{-1}$ is studied in \cite{TSh:17}:
\begin{theorem}\label{t:T}
For the 1d RBM of (\ref{Jb})  with $ N=C_*W^2$, we have
\begin{equation*}
\lim\limits_{n\to\infty}
D_2^{-1}\mathcal{R}_0\Big(E+\dfrac{\xi}{N\rho(E)}, E-\dfrac{\xi}{N\rho(E)}\Big)
=(e^{-C\Delta_U-2\pi i\xi\hat\nu}f_0,f_0),
\end{equation*}
where $C=1/t^*C_*$ with $t^*$ of (\ref{tens_pr}), and the limit is uniform in $\xi$ varying in any compact subset of $\mathbb{R}$. Here $E\in (-2,2)$.

\end{theorem}


\section{Analysis of $\mathcal{R}_1$}\label{s:4}
In the case of $\mathcal{R}_1$ the transfer operator $\mathcal{K}_1$ of (\ref{2}) has the form
\begin{align}\label{rep_R_1}
\mathcal{K}_1=  A_1(x_1,x_2) A_2(y_1,y_2)\hat Q,\quad \hat Q:=\left(\begin{array}{cc}
1+L(\bar x,\bar y)/W^2&-1/W^2\\
-L(\bar x,\bar y)&1
\end{array}\right)
\end{align}
with some explicit function $L$ whose form is not important for us now.
Operators $A_{1,2}$ 
(the same as for $\mathcal{R}_0$) contain a large parameter $W$ in the exponent, hence  only $W^{-1/2}$-
neighbourhood of the stationary point gives the main contribution. The spectral analysis of  $A_1$ gives us
that
\begin{align}\notag
&A_{1}\sim e^{\xi g_+(E)/N}A_+,\quad A_{2}\sim A_+,\\
&A_+(x,y)=(2\pi)^{-1/2}W^2e^{-W^2(x-y)/2+c_+(x^2+y^2)/2},\quad c_+=1+a_+^{-2},\notag\\
& g_+(E)=(-E+i\sqrt{4-E^2})/{2}.
\label{g(E)}
\end{align}
Then since 
\begin{align}\label{tA.1}
&\lambda_{j}(A_+)=\Big(1+\dfrac{2\alpha_+}{W}+\dfrac{c_+}{W^2}\Big)^{-1/2-j},\\
\label{alpha}
& \alpha_+=\sqrt{\dfrac{c_+}{2}}\Big(1+\frac{c_+}{2W^2}\Big)^{1/2},
 \end{align}
 we obtain that the spectral gap for $A_{1,2}$ is of the order $W^{-1}\gg N^{-1}$, hence one could expect that $A_1^{N-1}$ converges in the strong vector
 topology to the projection
 \[ A_{1,2}^{N-1}\to \lambda_0^{N-1}(A_1)\psi_0\otimes\psi_0^*\]
 where
 \[A_1\psi_0=\lambda_0(A_1)\psi_0,\quad A_1^*\psi_0^*=\overline{\lambda_0(A_1)}\psi_0^*.\]
The entry $Q_{12}$ here is  small hence
 the main order of our operator contains the Jordan cell. A simple computation shows that if we just replace in (\ref{rep_R_1})
$A_{1,2}$ by $A_+$ and $Q_{12}$ by $0$, then the answer will be wrong. Hence one should apply more refine analysis. An important point of such
analysis is an application of the "gauge"  transformation of $\mathcal{K}_1$ with  matrix $T$
\begin{align}\label{gauge}
&\mathcal{K}_1\to \mathcal{K}_{1T}=T\mathcal{K}_1T^{-1}=A_1A_2\hat S,\quad \hat S=T\hat Q T^{-1};\\
&T=\left(\begin{array}{cc}0&W^{-1/2}\\W^{1/2}&0\end{array}\right),
\quad \hat S=\left(\begin{array}{cc}1&-L/W\\-1/W&1+L/W^2\end{array}\right).\notag
\end{align}
With this transformation  it can be shown that for any $W$  
\[\lambda_0(\mathcal{K}_{1T})=e^{\xi g_+(E)/N}(1+O(n^{-2})),\quad |\lambda_1(\mathcal{K}_{1T})|\le 1-c/W,\; \quad c>0.\]

Hence for any $ 1\ll W\ll N$ we get that $(\mathcal{K}_{1T})^{N-1}$ converges in the strong vector topology to the projection (non-orthogonal) on the eigenvector, corresponding to $\lambda_0(\mathcal{K}_{1T})$. This gives
\begin{theorem}\label{t:0} 
Let $H$ be 1d Gaussian RBM defined in (\ref{Jb}) with $N\ge C_0W\log W$, and let $|E|\le 4\sqrt{2}/3\approx 1.88$. 
\begin{align*}
\mathcal{R}_1(E+\xi/N,E)\to e^{\xi g(E)},\quad\Big|\frac{\partial}{\partial \xi}\mathcal{R}_1(E+\xi/N,E)\Big|_{\xi=0}-g_+(E)\Big|\le C/W.
\end{align*}

The second relation implies that
\begin{equation} \label{rho_bound}
|\bar \rho_N(E)-\rho(E)|\le C/W,
\end{equation}
where $\bar \rho_N(E)=R_1(E)$ is the first correlation function, and $\rho(E)$ is defined in (\ref{rho_sc}).
\end{theorem}
\begin{remark} 
 The statement is expected to be true for all $|E|<2$.
The condition $|E|\le 4\sqrt{2}/3\approx 1.88$ is technical, and  it can be removed by the proper deformation of the integration contour in the
integral representation. 
\end{remark}
 Theorem \ref{t:0}  yields, in particular, that for $g_N(E+i\varepsilon)$ (the Stieltjes  transform of the first correlation function $\bar \rho_N(E)$) 
 and $g(E+i\varepsilon)$  (the Stieltjes  transform of  $ \rho(E)$) we have
\begin{equation}\label{loc_sc}
|\bar g_N(E+i\varepsilon)-g(E+i\varepsilon)|\le C/W
\end{equation}
uniformly in any arbitrary small $\varepsilon\ge 0$. As it was mentioned above, similar asymptotics (with correction $C/W^2$) for RBM of (\ref{Jb}) in 3d
was obtained in \cite{DPS:02} and in 2d was obtained in \cite{DL:16} (by the same techniques), however their method cannot be directly applied to 1d case since it  essentially uses the Fourier analysis which is different in 1d. 
All other previous results about the density of states
for RBM deal with $\varepsilon\gg ~W^{-1}$ or bigger (for fixed $\varepsilon>0$ the asymptotics  (\ref{loc_sc}) follows from the results of \cite{BMP:91};
\cite{EK:11} gives (\ref{loc_sc}) with $\varepsilon\gg  W^{-1/3}$; \cite{S:11} yields (\ref{loc_sc}) for 1d RBM with Bernoulli elements distribution for
$\varepsilon\ge W^{-0.99}$, and \cite{EYY:10} proves similar to (\ref{loc_sc})  asymptotics with correction $1/(W\varepsilon)^{1/2}$ for $\varepsilon\gg  1/W$).
On the other hand, the methods of \cite{EK:11}, \cite{EYY:10} allow to control $N^{-1}\Tr (E+i\varepsilon-H_N)^{-1}$  and $(E+i\varepsilon-H_N)^{-1}_{xy}$ 
for $\varepsilon\gg  W^{-1}$ without expectation, which gives some information about the localization length. This cannot be obtained from Theorem \ref{t:0}, since it requires
estimates on $\mathbb{E}\{|(E+i\varepsilon-H_N)^{-1}_{xy}|^2\}$.

\section{Analysis of $\mathcal{R}_2$ for the block RBM}\label{s:5}

\subsection{Sigma-model approximation for $\mathcal{R}_2$ for the block RBM}
We start from the analysis of  so-called sigma-model approximation for the model (\ref{H}) -- (\ref{J_old}). Sigma-model approximation
 is often used by physicists to study a complicated statistical mechanics systems.  In such approximation spins take values in
some symmetric space ($\pm 1$ for Ising model, $S^1$ for the rotator, $S^2$ for the classical
Heisenberg model, etc.). It is expected that sigma-models have all the qualitative physics
of more complicated models with the same symmetry (for more details see, e.g., \cite{Sp:12}).  The sigma-model approximation 
for RBM was introduced by Efetov (see \cite{Ef}),
and the spins there are $4\times 4$ matrices with both complex and Grassmann entries (this approximation was studied in \cite{FM:91}, \cite{FM:94}).
 Let us mention also the paper  \cite{SpDZ}, where  the average conductance for 1d Efetov's sigma-model for RBM was
computed.

In the subsection we present rigorous results on the derivation of the sigma-model approximation for 1d RBM 
and the analysis of the model in the delocalization  regime. The results are published in \cite{SS:sigma}.

To derive a sigma-model approximation for the model (\ref{H}) -- (\ref{J_old}), we take $\alpha$ in (\ref{J_old})  $\alpha=\beta/W$, i.e. put
\begin{equation}\label{J}
J=1/W+\beta\Delta/W^2, \quad \beta>0,
\end{equation}
fix $\beta$ and $n$, and consider the limit $W\to\infty$, for the generalized correlation functions

\begin{align}\label{G_2}
\mathcal{R}_{Wn\beta}^{+-}(E,\varepsilon,\xi)&=\mathbf{E}\bigg\{\dfrac{\mdet(H-z_1)\mdet(H-\overline{z}_2)}
{\mdet(H-z_1')\mdet(H-\overline{z}_2^\prime)}\bigg\},\\ \notag
\mathcal{R}_{Wn\beta}^{++}(E,\varepsilon,\xi)&=\mathbf{E}\bigg\{\dfrac{\mdet(H-z_1)\mdet(H-z_2)}
{\mdet(H-z_1')\mdet(H-z_2^\prime)}\bigg\}
\end{align}
for $\xi=(\xi_1,\xi_2,\xi_1^\prime,\xi_2^\prime)$.

\begin{theorem}\label{thm:sigma_mod} 
Given $\mathcal{R}_{Wn\beta}^{+-}$ of (\ref{G_2}) ,(\ref{H}) and  (\ref{J}),
with any dimension $d$,  any fixed $\beta$, $|\Lambda|$, $\varepsilon>0$, and 
 $\xi=(\xi_1,\bar\xi_2, \xi_1',\bar\xi_2')\in\mathbb{C}^4$ 
 ($|\Im\xi_j|<\varepsilon\cdot \rho(E)/2$) we have,   as $W\to\infty$:
\begin{align}\label{sigma-mod}
&\mathcal{R}_{Wn\beta}^{+-}(E,\varepsilon, \xi)\to\mathcal{R}_{n\beta}^{+-}(E,\varepsilon, \xi),\quad
\frac{\partial^2\mathcal{R}_{Wn\beta}^{+-}}{\partial\xi_1'\partial\xi_2'}(E,\varepsilon,\xi)\to
\frac{\partial^2\mathcal{R}_{n\beta}^{+-}}{\partial\xi_1'\partial\xi_2'}(E,\varepsilon, \xi),
\\ \hbox{where}\quad&\mathcal{R}_{n\beta}^{+-}(E,\varepsilon, \xi)=
C_{E,\xi}\int \exp\Big\{\dfrac{\tilde\beta}{4}\sum\Str Q_jQ_{j-1}-\dfrac{c_0}{2|\Lambda|}\sum \Str Q_j\Lambda_{\xi, \varepsilon}\Big\} d Q,
\notag\end{align}
  $\tilde\beta=(2\pi\rho(E))^2\beta$, $U_j\in \mathring{U}(2)$, $S_j\in \mathring{U}(1,1)=U(1,1)/U(1)\times U(1)$,
\begin{equation*}
C_{E,\xi}=e^{E(\xi_1+\xi_2-\xi_1'-\xi_2')/2\rho(E)},\quad \rho(E)=(2\pi)^{-1}\sqrt{4-E^2},
\end{equation*}
and $Q_j$ are $4\times 4$ supermatrices with commuting
diagonal and anticommuting off-diagonal $2\times 2$ blocks
\begin{align}\label{Q}
 Q_j=\left(\begin{array}{cc}
U_j^*&0\\
0&S_j^{-1}
\end{array}\right)\left(\begin{array}{cc}
(I+2\hat\rho_j\hat\tau_j)L&2\hat\tau_j\\
2\hat\rho_j&-(I-2\hat\rho_j\hat\tau_j)L
\end{array}\right)\left(\begin{array}{cc}
U_j&0\\
0&S_j
\end{array}\right),
\end{align}
\begin{align*}
d Q=\prod d Q_j,\quad d Q_j=(1-2n_{j,1}n_{j,2})\, d \rho_{j,1}d \tau_{j,1}\,d \rho_{j,2}d \tau_{j,2}\, d U_j\,d S_j
\end{align*}
with
\begin{align*}
&n_{j,1}=\rho_{j,1}\tau_{j,1},
\quad n_{j,2}=\rho_{j,2}\tau_{j,2},\\ \notag 
&\hat\rho_j=\mathrm{diag}\{\rho_{j1},\rho_{j2}\},\quad \hat\tau_j=\mathrm{diag}\{\tau_{j1},\rho_{j2}\},\quad L=\mathrm{diag}\{1,-1\}
\end{align*} 
Here $\rho_{j,l}$, $\tau_{j,l}$, $l=1,2$ are anticommuting Grassmann variables,
\[
\Str \left(\begin{array}{cc}
A&\sigma\\
\eta&B
\end{array}\right)=\Tr A-\Tr B,
\]
and 
\begin{align*}
\Lambda_{\xi,\varepsilon}=\mathrm{diag}\,\{\varepsilon-i\xi_1/\rho(E),-\varepsilon-i\xi_2/\rho(E),\varepsilon-i\xi_1'/\rho(E),-\varepsilon-i\xi_2'/\rho(E)\}.
\end{align*}\end{theorem}

\begin{theorem}\label{t:2}
Given $\mathcal{R}_{Wn\beta}^{++}$ of (\ref{G_2}) ,(\ref{H}) and  (\ref{J}),
with any dimension $d$,  any fixed $\beta$, $|\Lambda|$, $\varepsilon>0$, and 
 $\xi=(\xi_1,\xi_2, \xi_1',\xi_2')\in\mathbb{C}^4$ 
 ($|\Im\xi_j|<\varepsilon\cdot\rho(E)/2$) we have,   as $W\to\infty$:
\begin{align}\label{G++_lim}
&\mathcal{R}_{Wn\beta}^{++}(E,\varepsilon,\xi)\to e^{{ia_+}(\xi_1'+\xi_2'-\xi_1-\xi_2)/{\rho(E)}},\\
&\frac{\partial^2\mathcal{R}_{Wn\beta}^{++}}{\partial\xi_1'\partial\xi_2'}(E,\varepsilon,\xi)\to-a_+^2/\rho^2(E)\cdot e^{{ia_+}(\xi_1'+\xi_2'-\xi_1-\xi_2)/{\rho(E)}},
\qquad a_{+}=(iE+\sqrt{4-E^2})/{2}.
\notag\end{align}
\end{theorem}
Note that $Q_j^2=I$ for $Q_j$ of (\ref{Q}) and so the integral in the r.h.s of (\ref{sigma-mod}) is a sigma-model approximation similar to Efetov's one (see \cite{Ef}).

The kernel of the transfer operator for $\mathcal{R}_2^{(\sigma)}$  has a form
\begin{align*}
\mathcal{K}_2^{(\sigma)}=\hat F\hat Q \hat F
\end{align*}
where $\hat F$ and $\hat Q$ are $6\times 6$ matrix kernels, such that  $\hat F_{\mu\nu}$ are the operators
of the multiplication by some function of $U,S$  and $\hat Q_{\mu\nu}=\hat Q_{\mu\nu}(U_1U_2^*,S_1S_2^{-1})$ are
the "difference" operators.\\
After some asymptotic analysis $\mathcal{K}_2^{(\sigma)}$ and some "gauge" transformation  similar to (\ref{gauge})
we obtain that $T\mathcal{K}_2^{(\sigma)}T$ can be replaced by the $4\times 4$ "effective" matrix kernel
\begin{align}\label{eff}
&\hskip4cm    T\mathcal{K}_2^{(\sigma)}T\sim\tilde F\hat K_0 \tilde F,\\
&\quad \hat K_0=\left(\begin{array}{cccc}K&\tilde K_1&\tilde K_2&\tilde K_3\\
0&K&0&\tilde K_2\\0&0&K&\tilde K_1\\0&0&0&K\end{array}\right),\quad \tilde F=F\left(\begin{array}{cccc}1&\tilde F_1 &\tilde F_2&\tilde F_1 \tilde F_2\\
0&1&0&\tilde F_2\\0&0&1&\tilde F_1\\0&0&0&1\end{array}\right)
\notag\end{align}
where $K=K_U\otimes K_S$
\[ K_U(U_1,U_2)\sim \beta e^{-\beta|(U_1U_2^*)_{12}|^2},\,\,K_S(S_1,S_2)\sim \beta e^{-\beta|(S_1S_2^{-1})_{12}|^2},
\] 
$\tilde K_i=\tilde K_i(U_1U^*_2;S_1S_2^{-1})$,   $F$ is an operator of multiplication by $e^{\varphi(U,S)/2n}$,  and $\tilde F_{1,2}$ 
are  operators of multiplication by $n^{-1}\varphi_{1,2}(U,S)$ with some specific $\varphi$, $\varphi_1$ and $\varphi_2$. An important feature of $\tilde K_i$ that they satisfy the operator bound
\[|\tilde K_i|\le C\beta^{-1}(\Delta_U+\Delta_S)\]
where $\Delta_U,\Delta_S$ are the Laplace operator on the correspondent groups (see e.g. (\ref{Lapl_U}) for the definition of $\Delta_U$).
The bounds imply that for sufficiently smooth function $f$ $\tilde K_i f\sim \beta^{-1}$.

Similarly to Section 3 the idea is to show that in the regime $\beta\gg  n$ 
\[\tilde F\hat K_0 \tilde F\sim \tilde F^2\]
Then we get
\begin{align*}
\mathcal{R}_{n\tilde{\beta}}^{+-}(E,\varepsilon,\xi)&=\frac{C^*_E}{2\pi i}\oint_{\omega_A}z^{n-1}
(\widehat G_0(z)\widehat f,\widehat g)dz+o(1)=C^*_E(\widehat F^{2n-2}\widehat f,\widehat g)+o(1)\\
&=C^*_E\int \big(4n^2F_1F_2-2) F^{2n}dUdS+o(1), \end{align*}
where
\begin{align}\label{C^*}
 C^*_E=e^{-g_+(E)(\xi_1+\xi_1'-\xi_2-\xi_2')/\rho(E)},\quad g_{+}(E)=(-E+i\sqrt{4-E^2})/2.
\end{align}

This relation allows us to prove 
\begin{theorem}\label{t:1} If $n,\beta\to\infty$ in such a way that $\beta>Cn\log^2n$, then for any fixed $\varepsilon>0$ and 
 $\xi=(\xi_1,\xi_2, \xi_1',\xi_2')\in\mathbb{C}^4$ ($|\Im\xi_j|<\varepsilon\cdot\rho(E)/2$) we have
\begin{align}\label{t1.1}
\mathcal{R}^{+-}_{n\beta}\to &C^*_E\Big(\frac{\delta_1\delta_2}{\alpha_1\alpha_2}(e^{2c_0\alpha_1}-1)
-\frac{\delta_1+\delta_2}{\alpha_2}e^{2c_0\alpha_1}+e^{2c_0\alpha_1}\frac{\alpha_1}{\alpha_2}\Big),\\
\label{alp}
\hbox{where}\quad&\alpha_1=\varepsilon-{i(\xi_1-\xi_2)}/{2\rho(E)},\quad \alpha_2=\varepsilon-{i(\xi_1'-\xi_2')}/{2\rho(E)},\\
&\delta_1={i(\xi_1'-\xi_1)}/{2\rho(E)},\quad\delta_2={i(\xi_2-\xi_2')}/{2\rho(E)},
\notag\end{align}
and $ C^*_E$ is defined in (\ref{C^*}).
\end{theorem}
Theorem \ref{t:1} combined with Theorem \ref{t:2}
gives the GUE type behaviour  for the spectral correlation function:
\begin{theorem}\label{t:cor}
In the dimension $d=1$ the behaviour of the sigma-model approximation of the second order correlation function (\ref{G_2}) of (\ref{H}), (\ref{J}), as $\beta\gg  n$,
in the bulk of the spectrum coincides with those for the GUE. More precisely, if $\Lambda=[1,n]\cap \mathbb{Z}$ and $H_N$, $N=Wn$ are  matrices (\ref{H}) with $J$ of (\ref{J}), then 
for any $|E|<\sqrt{2}$
(\ref{Un}) holds
in the limit first $W\to\infty$, and then $\beta, n\to\infty$, $\beta\ge Cn \log^2 n$.
\end{theorem}

\subsection{Analysis of $\mathcal{R}_2$ for  block RBM of (\ref{H})-(\ref{J_old})}
As it was mentioned in Section \ref{s:2}  in the case of $\mathcal{R}_2$ the transfer operator $\mathcal{K}_2$ is a $70\times 70$ matrices whose
entries depend on 8 spacial variables $x_{1},x_{2},y_{1},y_{2};x_{1}',x_{2}',y_{1}',y_{2}'\in \mathbb{R}$, two unitary $2\times 2$ matrix $U, U'$,
 and two hyperbolic $2\times 2$ matrix $S,S'$,
which acts in  the direct sum of 70 Hilbert spaces
$ L_2(\mathbb{R}^4)\otimes L_2(\mathring{U}(2),dU)\otimes L_2(\mathring{U}(1,1),dS)$,
 where $dU, dS$ are integrations with respect to the
corresponding Haar measures. In general the analysis of such operator is a very involved problem, unless there is a possibility to 
take into account some special features of the matrix kernel and to reduce it (in the sense of Definition \ref{def:1}) by some matrix kernel
of smaller dimensionality.

In the case of $\mathcal{K}_2$ the first observation is that it can be  factorised as
\begin{align*}
\mathcal{K}_2=\hat F\hat Q\hat A \hat F
\end{align*}
where $\hat F$, $\hat Q$ and $\hat A$ are $70\times 70$ matrix kernels, such that $\hat F_{\mu\nu}$ are the operators
of  multiplication by some function of $U,S$,  
\begin{align*}
&\hat Q_{\mu\nu}=K_{U}K_{S}Q_{\mu\nu}(U(U')^*;S(S')^{-1}), \\
&K_{U}=\alpha t We^{-\alpha Wt|(U(U')^*)_{12}|^2 },\quad 
K_{S}=\alpha \tilde t We^{-\alpha W\tilde t|(S(S')^{-1})_{12}|^2 },
\end{align*}
with $t$, $\tilde t$ defined similarly to (\ref{t}) and functions $Q_{\mu\nu}$ which do not depend on $W$, and
\begin{align*}
\hat A_{\mu\nu}= A_1(x_1,x_1')A_2(y_1,y_1')A_3(x_2,x_2')A_4(y_2,y'_2)\mathcal{A}_{\mu,\nu}(\bar x,\bar x',\bar y,\bar y')
\end{align*}
with $A_{1,2,3,4}$ being a scalar kernels 
 similar to that  for $\mathcal{R}_0$ (see (\ref{A})) and functions $A_{\mu\nu}$ which do not depend on $W$.
 It is straightforward to prove  that  only $W^{-1/2}\log W$ -neighbourhoods of some stationary points in $\mathbb{R}^8$ give essential contributions. 
 Further analysis shows  that  after some "gauge" transformation similar to (\ref{gauge}) $T\mathcal{K}_2T^{-1}$ can be replaced
   (in the sense of Definition \ref{def:1}) by $4\times 4$ effective kernel of the form similar to (\ref{eff}).

Remark that the analysis justifies the physics conjecture  that the behaviour of the "generalized" correlation function $\mathcal{R}_2$ for the model  (\ref{H}) -- (\ref{J_old})
and of its sigma-model approximation  $\mathcal{R}_2^{\sigma}$ of are very similar.

As a result we obtain (cf with Theorem \ref{t:cor})
\begin{theorem}\label{t:bbrm}
In the dimension $d=1$ the behaviour of  the second order correlation function (\ref{cor=det}) of the model (\ref{H}) -- (\ref{J_old}), as $W\gg  n$,
in the bulk of the spectrum coincides with those for the GUE. More precisely, 
for any $|E|<\sqrt{2}$ (\ref{Un}) holds
in the limit  $W,n\to \infty$  with $W/\log^2W>Cn$.
\end{theorem}
The theorem is the main result of  the paper \cite{SS:deloc}.

\end{document}